\documentclass{article}

\usepackage[utf8]{inputenc} % allow utf-8 input
\usepackage[T1]{fontenc}    % use 8-bit T1 fonts
\usepackage{hyperref}       % hyperlinks
\usepackage{url}            % simple URL typesetting
\usepackage{booktabs}       % professional-quality tables
\usepackage{amsfonts}       % blackboard math symbols
\usepackage{amsmath,amsfonts,amsthm,dsfont,amssymb,amsxtra}
\usepackage{setspace,graphicx,color}
\newtheorem{theorem}{Theorem}

\newtheorem{proposition}[theorem]{Proposition}
\newtheorem{remark}[theorem]{Remark}
\def\mean{\mathbb{E}}
\def\R{\mathbb{R}}
\def\var{\mathbb{V}}
\def\suml{\sum_{\ell=1}^L}

\title{Architectures of epidemic models: accommodating constraints from empirical and clinical data}

\author{
  G.~Turinici\thanks{Webpage: https://www.turinici.com} \\
 CEREMADE\\
  University Paris Dauphine\\
  Paris 75016, France \\
  \texttt{gabriel.turinici@dauphine.fr} \\
}

\begin{document}
\maketitle

\begin{abstract}
Deterministic compartmental models have been used extensively in modeling epidemic propagation.
These models are required to fit available data and numerical procedures are often 
implemented to this end. But not every model architecture is able to fit the data because the structure of the 
model imposes hard constraints on the solutions. We investigate in this work two such situations: first the distribution of transition times from a compartment to another may impose a 
variable number of intermediary states; secondly, a non-linear relationship between 
time-dependent measures of compartments sizes may indicate the need for structurations (i.e., considering several groups of individuals of heterogeneous characteristics).
\end{abstract}

% keywords can be removed
%\keywords{epidemic model \and SI(N)R epidemic model \and SIR model \and mathematical epidemiology \and compartmental epidemic model \and hypo-exponential distribution}

\section{Introduction} \label{sec:introduction}

The recent COVID-19 epidemic stimulated to a large extent the research into the domain of mathematical epidemiology, prompted for instance by the hope to obtain faithful predictions of the ongoing pandemic unfold and assess pharmaceutical and non-pharmaceutical epidemic control interventions. 

One of the most easy to use frameworks 
is that of the deterministic compartmental models 
 \cite{diekmann2000mathematical,MR3409181,hethcote1987epidemiological,anderson1992infectious,murray_mathematical_2007}
(originating from the well known SIR model~\cite{kermack_contribution_1927} of Kermack and McKendrick)
that partitions all individuals in a population into several classes and prescribes a set of differential equations to model the evolution of the number of individuals in each class; in this set we can include  the SIR model and its variants  like SEIR, $SI^nR$, etc.) or more involved proposals \cite{ng_double_2003,turinici_danchin_immunity,structuration_ref,danchin_new_2020, mfgcovidemma20, heterogeneitycovid20, oriane_ade_covid20}.

In order to be effective, these models, that involve multiple parameters, have to be calibrated to available data, that is, to find the precise values of the parameters such that the model fits the observed data. 
Sometimes the available data can be in contradiction with the model and thus the model has to be dropped or evolved.

The goal of this paper is to discuss how can one design the compartmental model in order to have the right architecture and thus to eventually have a good calibration quality.

We explore two situation: first for the case of $S(E)I^n R$ models 
where a trick \cite{lloyd2001} can help deal with non-exponential passage distributions and secondly the situation of models where time-dependent measures of class sizes are available. 

\section{Basic model descriptions and notations}
\label{sec:notations}

\begin{figure}
\centering
\includegraphics[width=0.95\textwidth]{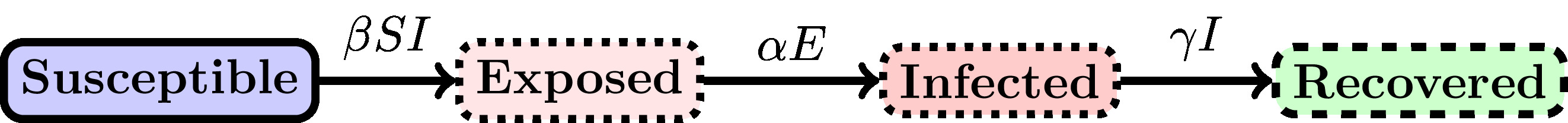}  \caption{Schematic evolution in the SEIR model.}
\label{fig:seir}
\end{figure}

We consider here the situation of a SEIR  ({\it Susceptible}, {\it Exposed}, {\it Infected}, {\it Recovered}) 
model~\cite{anderson1992infectious}  that is illustrated in figure~\ref{fig:seir} 
and whose mathematical transcription is: 
\begin{align}
&\frac{dS(t)}{dt}=-\,\beta\,S(t)\,\frac{I(t)}{N}\,,\label{eq:SEIR-S}\\
&\frac{dE\frac{I(t)}{N}}{dt}=\beta\,S\frac{I(t)}{N}-\gamma_E\,E(t)\,,\label{eq:SEIR-E}\\
&\frac{dI(t)}{dt}=\gamma_E\,E(t)-\,\gamma_I\,I(t)\,,\label{eq:SEIR-I}\\
&\frac{dR(t)}{dt}=\gamma_I\,I\,,\label{eq:SEIR-R} \\
& N = S(t)+E(t)+I(t)+R(t).
\end{align}

The SEIR model is relevant and has been used in many practical situations; it can also be seen as the average dynamics of individuals that follow a Markov chain dynamics between states 
 {\it Susceptible}, {\it Exposed}, {\it Infected}, and {\it Recovered}. For instance, to first order in $\Delta t$, the probability to go from "Susceptible" to "Exposed" during 
 $[t,t+\Delta t]$ is $\frac{\beta I(t)}{N} \Delta t$, from
 "Exposed" to "Infected" is $\alpha \Delta t$ and from "Infected" to "Recovered" is $\gamma \Delta t$. 

\begin{remark}
Similar equations arise when modelling the intra-host infection dynamics, see \cite{nowak_virus_2000_book,wodarz_killer_2007_book} for a classical introduction and \cite{oriane_ade_covid20} for use in the context of COVID-19. 
All results of this paper apply with straightforward adaptations.
\end{remark}
 
 \section{Non-exponential passage times between compartments: need for intermediary states} \label{sec:hypoexponential}
 %Models that accommodate 
 Let us focus on the "Infected" $\to$ "Recovered" transition. The time that an individual spends in the class $I$ is an exponential random variable of parameter $\gamma$; such a distribution can be empirically observed and unfortunately it happens often that the observed distribution  is not exponential but rather more close to a gamma distribution (see \cite{distributions_covid,distribution_covid2} for COVID-19 related information). In order to respect this constraint, proposals have been documented in the literature (see \cite{lloyd2001} for an entry point to this literature)  
that employ the
\textit{linear chain trick}; it consists in replacing a single infectious stage with $K$ identical exponentially distributed sub-stages, each having a mean period $1/(K \gamma)$. 
The historical approach is to divide the  infected population into $K$ identical stages $I^1, I^2, ..., I^K$ to create the so-called $SI^{K}R$ model and we denote the total infected population by $I=\sum_{i=1}^K I_{i}$. 

We depart from this setting by considering a more general model with $K$ stages that have each a different parameter $\gamma_k$.

In this section we ask the following question: can any continuous distribution on $[0,\infty[$ be represented using the
linear chain trick ? Or, put it otherwise, are there any constraints on some observed distribution of residence times in the {\it Infected} state that is the sum of $K$ independent exponential random variables of parameters 
$\gamma_1$, ..., $\gamma_K$ ? Note that such a distribution is called {\it hypoexponential distribution} and has already been used in mathematical biology, see \cite{hypoexpdistrib1,yates_multi-stage_2017}.

The continuous-time transcription of the model's equations is:
\begin{align}\label{sys}
\begin{cases}
\frac{dS(t)}{dt}=-\beta S(t) \frac{I(t)}{N},\\
\frac{dE(t)}{dt}=\beta S(t) \frac{I(t)}{N} - \gamma_E E(t),\\
\frac{d I_1(t)}{dt}= \gamma_E E(t) -\gamma_1 I_1(t),\\
\frac{dI_2(t)}{dt}=\gamma_K I_1(t)-\gamma_2 I_2(t),\\
 ... \\
\frac{dI_K(t)}{dt}=\gamma_{K-1} I_{K-1}(t)-\gamma_K I_K(t) ,\\
\frac{dR(t)}{dt}=\gamma_K I_K(t), \\
N = S(t)+E(t) + \sum_{k=1}^K I_k(t) + R(t),
\end{cases}
\end{align}
here $S$, $E$, $I$ and $R$ are the numbers of individuals in the {\it Susceptible},
{\it Exposed}, {\it Infected} and respectively {\it Recovered} classes.
The initial distribution of individuals at time $0$ is $(S(0)$, $E(0)$, $I_1(0)$, ..., $I_K(0)$, $R(0))$, which is supposed to be known (otherwise is numerically fitted to data). Note that the system $\eqref{sys}$ possesses a conservation law, i.e. $S(t)+E(t)+\sum_k I_k(t)+R(t)=\text{constant}$, for all $t \geq 0$.

Here $1/\gamma_k$ is the average number of days a patient spends in class $I_k$. We prove the following

\begin{proposition}
\begin{enumerate}
\item \label{item:hypoexp_estimation}
Let $X$ be a positive random variable. Then if 
\begin{equation}
\var(X) > \mean[X]^2,        
\end{equation}
then $X$ is not a hypo-exponential variable, i.e., cannot be represented as a sum  of independent exponentially distributed random variables $X_k$:
\begin{equation}
X=\sum_{k=1}^K X_k, \ X_k \simeq \text{Exp}(\gamma_k), \ \gamma_k >0, \ \forall k=1,...,K.
\label{eq:hypoexpsum}
\end{equation}
\item \label{item:representation_e_v}
On the contrary, given two strictly positive numbers $e,v > 0$ with $v \le e^2$
there exist $K$ and a hypo-exponential distribution satisfying \eqref{eq:hypoexpsum} 
such that $\mean[X]=e$  and $\var(X)=v$.
\item \label{item:hypoexp_number}
If $X$ is a hypoexponential variable that can be represented as in equation \eqref{eq:hypoexpsum} then
$K  \ge \left\lfloor \frac{\mean[X]^2}{\var(X)}\right\rfloor$.
\end{enumerate}
Here $\left\lfloor \cdot\right\rfloor$ is the operator returning the largest integer smaller than the argument, and the operators 
$\mean[\cdot]$ and $\var[\cdot]$ are the mean and variance with respect to the law of $X$.
\end{proposition}

\begin{proof}
Suppose \eqref{eq:hypoexpsum} is true. Then $\mean[X] = \sum_{k=1}^K (1/\gamma_k) $ and 
$\var[X] = \sum_{k=1}^K (1/\gamma_k^2) $. 
Of course, since $\gamma_k \ge 0$,
$\mean[X]^2 = \left( \sum_{k=1}^K 1/\gamma_k \right)^2  \ge \sum_{k=1}^K (1/\gamma_k^2) = \var[X]$ which proves the point \ref{item:hypoexp_estimation}. On the other hand, using the Cauchy-Schwartz inequality
$\mean[X]^2 = \left( \sum_{k=1}^K 1/\gamma_k \right)^2  \le K \sum_{k=1}^K (1/\gamma_k^2) 
= K \var[X]$ which shows that $K \ge\mean[X]^2/\var[X]$ and proves the point \ref{item:hypoexp_number} of the proposition.
Denote now $e=\sum_{k=1}^K (1/\gamma_k)$ and let us enquire about the possible values of 
$v=\sum_{k=1}^K (1/\gamma_k)^2$; the previous considerations show that 
$v \in [e^2/K,e^2]$. The lowest bound is attained exactly when $\gamma_k$ are all equal; the upper bound is attained asymptotically when all $\gamma_k$ go to zero except one of them that
tends to $1/e$ (and it is obtained exactly when $K=1$). Taking any continuous deformation 
of $\gamma_k$ (as a $K$-tuple)
between these extreme values (while keeping $e$ constant) one can show that all values in the interval $v \in [e^2/K,e^2[$ are attained. Making now $K\to \infty$ and adding the 
situation $K=1$ we obtain that all $v \in ]0,e^2]$ are attained, hence the item
\ref{item:representation_e_v} of the proposition is also proved.
\end{proof}

\begin{remark}
The point \ref{item:hypoexp_number} informs on the minimal number of stages that are necessary to implement a given distribution; this is not related to the uniqueness (in many situations a given hypo-exponential distributed random variable can only be decomposed in an unique way as sum of independent exponential variables) but rather as an indication that  more the distribution departs from the exponential situation
(which corresponds to$\frac{{\mathbb V}(X)}{{\mathbb E}[X]^2}=1$) more stages are necessary to describe it.
Reciprocally, the item \ref{item:representation_e_v} is not an existence result but only 
informing that comparing then mean and variance cannot bring additional information on whether some distribution can be represented as hypo-exponential or not. One may 
 can investigate in the same vein the domain of possible values for the triple
 (mean, variance, skewness) 
(skewness has an explicit simple form $2 \left( \sum_{k=1}^K 1/\gamma_k^3 \right) / \left( \sum_{k=1}^K 1/\gamma_k^2 \right)^{3/2} $).
\end{remark}

 \section{Nonlinear relationships between sizes of the compartments: the need for structuration}
\label{sec:structuration}

Let us now consider a different view of the passage from one state to another. We will take as model the passage from $I$ to $R$ in figure \ref{fig:seir} although other situations are also relevant (we will give a different example below). 

The discussion in this section is independent of that 
in section \ref{sec:hypoexponential}. If the 
\eqref{eq:SEIR-R} holds true with a constant $\gamma_I$ then of course,
$R(t)=R(0) + \int_0^t \gamma_I I(u) du$. In particular, considering $R(0)=0$ (otherwise the initial value 
has to be subtracted), the functions $R(\cdot)$ and $t \mapsto \int_0^t I(u) du$
will be linearly dependent; supposing empirical data on those two functions is available (usually the public statistics communicate such information), this requirement can be checked irrespective of 
that fact that the constant $\gamma_I$ is known or not (i.e., before any numerical fitting procedure).

We will take a general notation to designate as $A$ the source state and $B$ the 
secondary state, i.e. the flow of individuals is from the state "A" to state "B" at rate $\gamma$, i.e., on average, an individual stays in state "A" for $1/\gamma$ days.
As an example "A" can be the the hospitalized state and 'B'  the deceased (see next section). The 
mathematical formula relating the two is 
\begin{equation}
    \frac{d}{dt} B(t) = \gamma A(t). \label{eq:ab}
\end{equation}

\subsection{Illustration}

As an illustration let us consider the situation for COVID-19 where two figures are most often recovered in the public health authorities statistics, namely the number of hospitalizations and the death toll. These figures are indeed of higher statistical quality than the number of infections which is depending on the definition and of further considerations (e.g., accounting of asymptomatics, etc.). We will consider moreover a model where hospitalization is a state of the Markov chain (sometimes labeled "severe cases") linked directly to "deaths" state (counting deceased).

\begin{figure}
\centering
\includegraphics[width=0.95\textwidth]{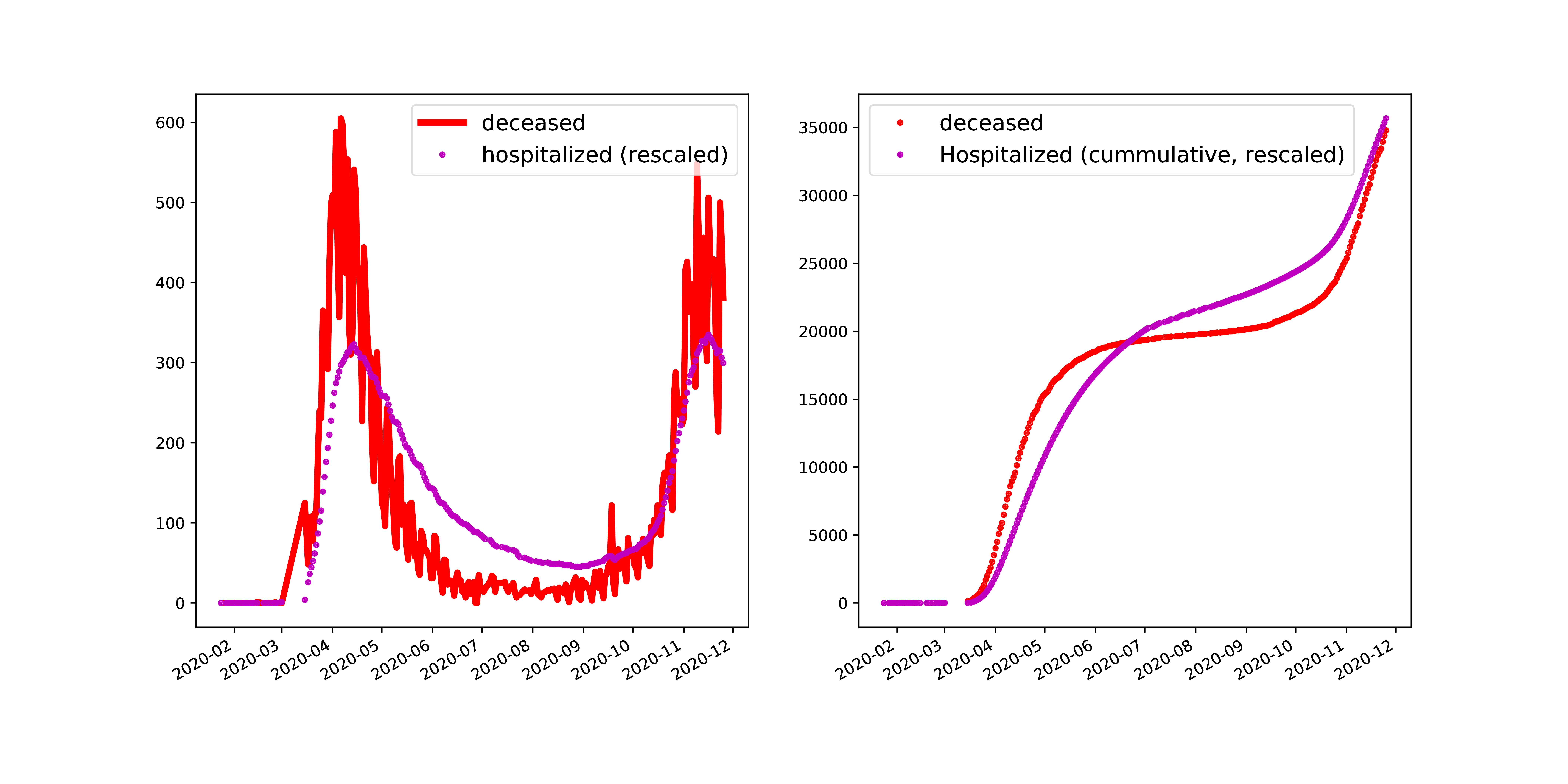}  
\caption{Empirical data for France concerning the COVID-19 hospitalizations and deceased (source "Sant\'e Publique France").
We use the notations "A" = hospitalized, "B"= deceased. 
{\bf Left:} plot of the numbers of hospitalized (rescaled arbitrarily for graphical convenience) (i.e., $A(t)$)
and the daily deaths (i.e. with our notations $B(t+1)-B(t)$ as an approximation of
$\dot{B}(t)$) {\bf Right}. Cumulative versions of the above, i.e., 
$\int_0^t A(t)$ and $B(t)$. In both cases, no linear relationship  between the two exists.
}
\label{fig:empirical_hosp_dc}
\end{figure}

The data is plotted in figure \ref{fig:empirical_hosp_dc}. Although both curves are related, there is no linear relationship between them and the differences are both qualitative and quantitative.
%take the example from \cite{turinici_danchin_immunity} 
%(see \cite{structuration_ref} for additional epidemic models used in COVID-19 practice).

\subsection{Structuring the model: theoretical constraints}

The fact that the linear relationship expected from equation \eqref{eq:ab} is not true in practice means that, in full rigor, the model cannot be used. It has to be fixed. One way to address this issue is to {\it structure} the model i.e., to recognize that not all individuals are alike and the average time 
of $1/\bar{\gamma}$ days 
spent in state 'A' does not mean everybody spends exactly this amount of time in that compartment.

We consider then a structuration by this $\gamma$ parameter, that is, we consider that each individual
has its own $\gamma$ belonging to some set of possible values 
$\{\gamma_1, ..., \gamma_L \}$. 
This view can readily be extended to accommodate an infinite set of possible values $[0,\infty[$
and a probability measure defined on it. The equation  \eqref{eq:ab} is to be replaced, with obvious notations, by:
\begin{equation}
    \frac{d}{dt} B_\ell(t) = \gamma_\ell A_\ell(t), \ \ell=1,..., L. \label{eq:abk}
\end{equation}

What is less obvious is to understand to what correspond the empirical measured values; for instance, the 
number of daily hospitalisations would correspond to $A(t) = \suml A_\ell(t)$, the
cumulative death toll to $B(t) = \suml B_\ell(t)$; however the distribution of values of
$\gamma$ that is obtained is not the real $\gamma$ distribution but instead is the distribution conditioned on the fact that the final state at final time $T$ is "B". If we denote by $X(t)$ the state of an individual at time $t$, for instance the empirical variance of $\gamma$ (that can be read from 
the empirical clinical data) will be $\var[\gamma | X(T) = "B"]$\footnote{With a slight abuse of language we identify the random variable $\var[\gamma | X(T) = "B"]$ with its value 
on the set $\{X(T) = "B"\}$.}.

Accordingly, when we refer to mean or variance of $1/\gamma$ we will in fact refer to:
\begin{eqnarray}
& \ & 
\mean \left[\frac{1}{\gamma}\right] = \suml \frac{1}{\gamma_\ell} \frac{B_\ell(T)}{B(T)},
\\ & \ & 
\var\left[\frac{1}{\gamma}\right] = \suml \left( \frac{1}{\gamma_\ell} \right)^2 \frac{B_\ell(T)}{B(T)}
- \mean \left[\frac{1}{\gamma}\right] ^2. \label{eq:variance}
\end{eqnarray}

As in section \ref{sec:hypoexponential} we will prove a result involving those three measurable quantities: the time dependent data $A(t)$ and $B(t)$ and the conditional variance of $1/\gamma$.

\begin{proposition}
With the notations in \eqref{eq:variance} we have:
\begin{equation}
\var\left[\frac{1}{\gamma}\right]  
\ge 
\frac{ \min_{\xi \in \R} \|  A(t) - \xi \dot{B}(t) \|^2_{L^2(0,T)} }{B(T)^2}.
\label{eq:propstruct}
\end{equation}
\end{proposition}

\begin{proof}
We will replace in $\min_{\xi \in \R} \|  A(t) - \xi \dot{B}(t) \|^2_{L^2(0,T)} $
the minimum value by the average $\mean \left[\frac{1}{\gamma}\right]$ and write:
\begin{align}
\min_{\xi \in \R} \|  A(t) - \xi \dot{B}(t) \|^2_{L^2(0,T)}  =
\min_{\xi \in \R} \left\|  \suml A_\ell(t) - \xi \dot{B_\ell}(t) \right\|^2_{L^2(0,T)}  
\nonumber \\
= \min_{\xi \in \R} \left\|  \suml (1/\gamma_\ell) \dot{B_\ell}(t)- \xi \dot{B_\ell}(t) \right\|^2_{L^2(0,T)}  
\le \left\|  \suml \frac{1}{\gamma_\ell} \dot{B_\ell}(t)-  \mean\left[\frac{1}{\gamma}\right]   \dot{B_\ell}(t) \right\|^2_{L^2(0,T)}.
\end{align}
We continue with a upper bound:
\begin{align}
 \left\|  \suml \frac{1}{\gamma_\ell} \dot{B_\ell}(t)-  \mean\left[\frac{1}{\gamma}\right]   \dot{B_\ell}(t) \right\|^2_{L^2(0,T)}   =  \left\|  \suml \left( \frac{1}{\gamma_\ell} -  \mean\left[\frac{1}{\gamma}\right] \right)  \dot{B_\ell}(t) \right\|^2_{L^2(0,T)}  
\nonumber \\
= \int_0^T \left[
\suml \left( \frac{1}{\gamma_\ell} -  \mean\left[\frac{1}{\gamma}\right] \right)  \dot{B_\ell}(t)
\right]^2 dt
\nonumber \\
\le \left\{
\int_0^T \suml \left( \frac{1}{\gamma_\ell} -  \mean\left[\frac{1}{\gamma}\right] \right)^2  \dot{B_\ell}(t)  dt \right\} 
\cdot \left\{
\int_0^T 
\suml   \dot{B_\ell}(t) dt \right\}
\nonumber \\
= \left\{ \suml \left( \frac{1}{\gamma_\ell} -  \mean\left[\frac{1}{\gamma}\right] \right)^2 
B_\ell(T)
\right\} 
\cdot B(T) = \var\left[\frac{1}{\gamma}\right]  B(T)^2.
\end{align}

\end{proof}

\begin{remark}
The presence of the term $\min_{\xi \in \R} \|  A(t) - \xi \dot{B}(t) \|^2_{L^2(0,T)}$ in 
equation \eqref{eq:propstruct}
is to measure the difference with the un-structured situation (when it is null). In particular, when the term is non-null this means that no un-structured model can fit both observed data $A(t)$ and $B(t)$. The distribution of $\gamma$ has to satisfy \eqref{eq:propstruct}. This can be used to orient numerical procedures to fit the $\gamma$ parameters.
\end{remark}

\section{Concluding remarks}

We presented in this contribution how constraints coming from the empirical available data, 
(that  the epidemic models are fitted to reproduce before any further, e.g., previsional, use) impose constraints on
the model architecture: 
in the first case, in order to accommodate non-exponential passage times, the number of compartments 
needs to be increased and in the second case structuration by
some transmission parameters is necessary when there is no linear-relationship between some compartment related time-dependent functions. 
Additional constraints of this type can be found along the lines proposed in this paper but are left for future work. These considerations can help propose relevant models and are useful in the implementation of numerical procedures to find the model parameters and will ultimately impact the model quality.
\bibliographystyle{unsrt}  
%\bibliography{references} 

\begin{thebibliography}{10}

\bibitem{diekmann2000mathematical}
Odo Diekmann and Johan Andre~Peter Heesterbeek.
\newblock {\em Mathematical epidemiology of infectious diseases: model
  building, analysis and interpretation}, volume~5.
\newblock John Wiley \& Sons, 2000.

\bibitem{MR3409181}
Maia Martcheva.
\newblock {\em An introduction to mathematical epidemiology}, volume~61 of {\em
  Texts in Applied Mathematics}.
\newblock Springer, New York, 2015.

\bibitem{hethcote1987epidemiological}
Herbert~W Hethcote and James~W Van~Ark.
\newblock Epidemiological models for heterogeneous populations: proportionate
  mixing, parameter estimation, and immunization programs.
\newblock {\em Mathematical Biosciences}, 84(1):85--118, 1987.

\bibitem{anderson1992infectious}
Roy~M Anderson, B~Anderson, and Robert~M May.
\newblock {\em Infectious diseases of humans: dynamics and control}.
\newblock Oxford university press, 1992.

\bibitem{murray_mathematical_2007}
James~D. Murray.
\newblock {\em Mathematical {Biology}: {I}. {An} {Introduction}}.
\newblock Springer Science \& Business Media, June 2007.

\bibitem{kermack_contribution_1927}
W.~O. {Kermack} and A.~G. {McKendrick}.
\newblock {A contribution to the mathematical theory of epidemics}.
\newblock {\em {Proc. R. Soc. Lond., Ser. A}}, 115:700--721, 1927.

\bibitem{ng_double_2003}
Tuen~Wai Ng, Gabriel Turinici, and Antoine Danchin.
\newblock A double epidemic model for the {SARS} propagation.
\newblock {\em BMC Infectious Diseases}, 3(1):19, September 2003.

\bibitem{turinici_danchin_immunity}
Antoine Danchin and Gabriel Turinici.
\newblock Immunity after covid-19: Protection or sensitization?
\newblock {\em Mathematical Biosciences}, page 108499, 2020.

\bibitem{structuration_ref}
Giulia Giordano, Franco Blanchini, Raffaele Bruno, Patrizio Colaneri,
  Alessandro Di~Filippo, Angela Di~Matteo, and Marta Colaneri.
\newblock {Modelling the COVID-19 epidemic and implementation of
  population-wide interventions in Italy}.
\newblock {\em Nature Medicine}, 26:855 -- 860, 2020.

\bibitem{danchin_new_2020}
Antoine Danchin, Tuen Wai~Patrick Ng, and Gabriel Turinici.
\newblock A new transmission route for the propagation of the {SARS}-{CoV}-2
  coronavirus.
\newblock {\em medRxiv Doi:
  \href{https://doi.org/10.1101/2020.02.14.20022939}{10.1101/2020.02.14.20022939}},
  2020.

\bibitem{mfgcovidemma20}
{Elie, Romuald}, {Hubert, Emma}, and {Turinici, Gabriel}.
\newblock Contact rate epidemic control of covid-19: an equilibrium view.
\newblock {\em Math. Model. Nat. Phenom.}, 15:35, 2020.

\bibitem{heterogeneitycovid20}
{Dolbeault, Jean} and {Turinici, Gabriel}.
\newblock Heterogeneous social interactions and the covid-19 lockdown outcome
  in a multi-group seir model.
\newblock {\em Math. Model. Nat. Phenom.}, 15:36, 2020.

\bibitem{oriane_ade_covid20}
Antoine Danchin, Oriane Pagani-Azizi, Gabriel Turinici, and Ghozlane Yahiaoui.
\newblock Covid-19 adaptive humoral immunity models: non-neutralizing versus
  antibody-disease enhancement scenarios.
\newblock {\em medRxiv Doi:
  \href{https://doi.org/10.1101/2020.10.21.20216713}{10.1101/2020.10.21.20216713}},
  2020.

\bibitem{lloyd2001}
Alun~L. Lloyd.
\newblock Realistic distributions of infectious periods in epidemic models:
  Changing patterns of persistence and dynamics.
\newblock {\em Theoretical Population Biology}, 60(1):59 -- 71, 2001.

\bibitem{nowak_virus_2000_book}
Martin Nowak and Robert~M. May.
\newblock {\em Virus {Dynamics}: {Mathematical} {Principles} of {Immunology}
  and {Virology}}.
\newblock Oxford University Press, Oxford, New York, November 2000.

\bibitem{wodarz_killer_2007_book}
Dominik Wodarz.
\newblock {\em Killer cell dynamics: mathematical and computational approaches
  to immunology}, volume~32.
\newblock Springer, New York, 2007.
\newblock OCLC: 184908900.

\bibitem{distributions_covid}
Luca Ferretti, Chris Wymant, Michelle Kendall, Lele Zhao, Anel Nurtay, Lucie
  Abeler-D{\"o}rner, Michael Parker, David Bonsall, and Christophe Fraser.
\newblock {Quantifying SARS-CoV-2 transmission suggests epidemic control with
  digital contact tracing}.
\newblock {\em Science}, 368(6491), 2020.

\bibitem{distribution_covid2}
Natalie Linton, Tetsuro Kobayashi, Yichi Yang, Katsuma Hayashi, Andrei
  Akhmetzhanov, Sung-mok Jung, Baoyin Yuan, Ryo Kinoshita, and Hiroshi
  Nishiura.
\newblock Incubation period and other epidemiological characteristics of 2019
  novel coronavirus infections with right truncation: A statistical analysis of
  publicly available case data.
\newblock {\em Journal of Clinical Medicine}, 9(2):538, Feb 2020.

\bibitem{hypoexpdistrib1}
Enrico Gavagnin, Matthew~J. Ford, Richard~L. Mort, Tim Rogers, and Christian~A.
  Yates.
\newblock The invasion speed of cell migration models with realistic cell cycle
  time distributions.
\newblock {\em Journal of Theoretical Biology}, 481:91--99, November 2019.

\bibitem{yates_multi-stage_2017}
Christian~A. Yates, Matthew~J. Ford, and Richard~L. Mort.
\newblock A {Multi}-stage {Representation} of {Cell} {Proliferation} as a
  {Markov} {Process}.
\newblock {\em Bulletin of Mathematical Biology}, 79(12):2905--2928, December
  2017.

\end{thebibliography}

\end{document}